\documentclass[aps,preprint,showpacs,superscriptaddress,groupedaddress]{aastex}  
\usepackage{graphicx}  
\usepackage{dcolumn}   
\usepackage{bm}        
\usepackage{amssymb}   
\begin{document}


\title{Near-extremal black holes as initial conditions of long GRB-supernovae and probes of their gravitational wave emission}

\author{Maurice H.P.M. van Putten}
\affil{Astronomy and Space Science, Sejong University, 98 Gunja-Dong Gwangin-gu, Seoul 143-747, Korea\\}

\date{\today}


\begin{abstract}
Long gamma-ray bursts (GRBs) associated with supernovae and short GRBs with Extended Emission (SGRBEE) from mergers are probably powered
by black holes as a common inner engine, as their prompt GRB emission satisfies the same Amati correlation in the $E_{p,i}-E_{iso}$ plane. We introduce modified Bardeen equations to identify hyper-accretion driving newly formed black holes in core-collapse supernovae to near-extremal spin 
as a precursor to prompt GRB emission. Subsequent spin-down is observed in the BATSE catalog of long GRBs. Spin-down provides a natural unification of long durations associated with the lifetime of black hole spin for normal long GRBs and SGRBEEs, given the absence of major fallback matter in mergers. The results point to major emissions unseen in high frequency gravitational waves. A novel matched filtering method is described for LIGO-Virgo and KAGRA broadband probes of nearby core-collapse supernovae at essentially maximal sensitivity.
\end{abstract}

\maketitle


\section{Introduction}

\begin{figure}[h]
\centerline{\includegraphics[scale=0.35]{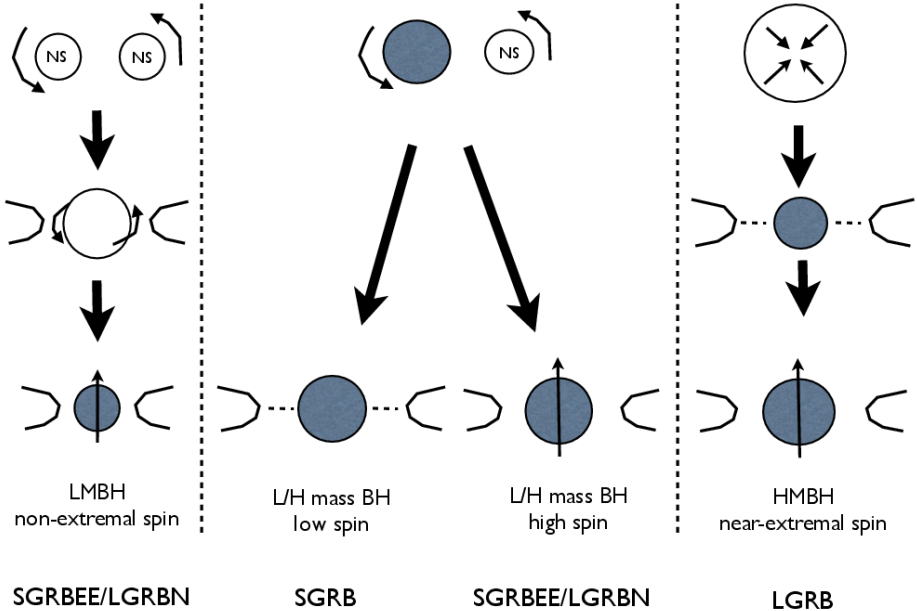}}
\caption{Overview of GRBs from rotating black holes originating in mergers of neutron stars with another neutron star (NS-NS, left) or stellar mass black hole (NS-BH, middle) and core-collapse supernovae (CC-SN, right). The GRB is produced by the resulting black hole-disk or torus system with low mass (LMBH) or high mass black hole (HMBH). NS-NS produces a low mass black hole with high but non-extremal spin following an intermediate phase of a super-massive near-extremal neutron star \citep{bai08,van13}.
The outcome of NS-BH mergers depends largely on the state of the BH in the progenitor binary, unless its mass and spin are low. Black hole formation in a CC-SN passes through hyper-accretion along a (modified) Bardeen trajectory, causing the black hole to surge to high mass and spin. Soft extended emission (EE) is identified with spin down of rapidly rotating black holes, producing SGRBEEs or LGRBs with no supernovae (LGRBN) from mergers involving rapidly rotating black holes and normal LGRBs from CC-SNe. This common origin of EEs finds is supported by a shared Amati-correlation.}
\label{fig:0}
\end{figure}

Long gamma-ray bursts (LGRBs) associated with supernovae \citep{woo93} and short GRBs with Extended Emission (SGRBEE) satisfy the same Amati correlation in the $E_{p,i}-E_{iso}$ plane, including long GRBs with no association to supernovae \citep[LGRBN;][]{ama02,van14a}. It points to black holes as a common inner engine to extended soft GRB emission. Long durations attributed to a state of suspended accretion \citep{van01b} hereby point to rapidly rotating black holes in both normal long GRBs and SGRBEEs \citep{van14a}. We hereby challenge the conventional association of prompt GRB emission from hyper-accretion onto black holes \citep[e.g.][]{woo93,pir04,kum08b}. 
Instead, we propose that the lifetime of black hole spin is a secular time scale common to normal long GRBs and SGRBEEs \citep{van03}. However, newly formed black holes in core-collapse supernovae do experience
a phase of hyper-accretion that generally leads to a surge in their mass and spin. Quite generally, therefore, these newly formed black holes
may evolve, after birth if kick velocities are sufficiently low, through possibly two phases of surge and spin-down. These phases offer possibly interesting windows to accompanying gravitational wave emission. Fig. \ref{fig:0} gives a schematic overview of GRBs from rotating black holes.

For LGRBs from CC-SNe, we here consider newly formed black holes as GRB inner engines evolving by hyper-accretion and interaction with surrounding high density matter. To this end, we  introduce a new system of modified Bardeen equations to describe evolution in the presence of ultra-relativistic black hole outflows. At canonical efficiencies, the black hole is hereby shown to proceed to a near-extremal state. Yet, BATSE light curves of long GRBs reveal spin-down of initially rapidly rotating black holes \citep{van12}. In the present work, we shall argue that hyper-accretion onto black holes serves as a {\em precursor} to the observed prompt GRB emission during the subsequent phase of spin down. Over the course of tens of seconds, spin down exhausts essentially all of the spin energy $E_r$ of the putative black hole, essentially unseen in the relatively limited energies observed in the true energy in gamma-rays $E_\gamma$ and kinetic energies in supernovae $E_{SN}$. 

To characterize inner engines powered by spin down of rotating black holes, we shall introduce calorimetric ratios of the energy output in various channels relative to the rotational energy $E_r$ of the black hole. A near-extremal Kerr black hole of mass $M$ satisfies
\begin{eqnarray}
E_{r}\simeq 6 \times 10^{54} \left(\frac{M}{10 M_\odot}\right)\,\mbox{ erg},
\end{eqnarray}
as follows from the Kerr metric of rotating black holes \citep{ker63}. It amply accounts for the true energy in gamma-rays
$E_\gamma$ of about $10^{51}$ erg \citep{fra01,ghi06,ghi13} and the kinetic energy in the most hyper-energetic GRB-supernovae. 
$E_\gamma$ is believed to derive from dissipation of $E_j$ in ultra-relativistic baryon-poor jets \citep[BPJ][]{lev93} at a canonical efficiencies $\epsilon$ implies a remarkably small ratio (Table 1)
\begin{eqnarray}
R_{j}=\frac{E_j}{E_r}\simeq 0.2\% \left(\frac{\epsilon}{0.16}\right)^{-1}.
\label{EQN_Rjr}
\end{eqnarray}
In exhausting $E_r$, an extremely minor fraction channels into $E_j$ (unless efficiencies are anomalously low). Note that $E_\gamma$ of SGRBEEs tends to be less than $E_\gamma$ of normal long GRBs, perhaps by different initial states at the onset of their soft extended emission. Table 1 further shows similarly small 
\begin{eqnarray}
R_{k} = \eta^{-1} \frac{E_{SN}}{E_r} \simeq \mbox{few}\,\%
\label{EQN_Rkr}
\end{eqnarray}
for the kinetic energy $E_{SN}$ in accompanying supernovae, here associated with ejection by magnetic winds at an efficiency $\eta$ \citep[][and references therein]{van11}. $E_{SN}$ points to additional baryon-rich disk winds which also cannot account for $E_r$. Exceptionally energetic events such as GRB 031203/SN2003lw and GRB 030329/SN2003dh require $E_r$ about an order of magnitude in excess of the maximal energy $E_c\simeq 3\times 10^{52}$ erg of a rapidly spinning neutron star.

In the absence of magnetic fields, accretion tends to settle down to thin disks \citep{sha73}. With magnetic fields, open outflows may be launched from rotating black holes and the accretion disk \citep{bla77,bla82,tho86,tho88}. A potentially different picture may result from dynamically strong magnetic fields, when the Alfv\'en speed reaches the hydrodynamic sound speed. Open outflows may then appear at enhanced accretion efficiency \citep{nar03,mck12,zam14}, possibly at extreme luminosities by intermittency \citep{van14b} by feedback onto matter at the Innermost Stable Circular Orbit (ISCO) mediated by horizon Maxwell stresses \citep{ruf75} through an inner torus magnetosphere \citep{van99a,van99b,van01b}. Conceivably, feedback brings accretion to a halt. Different accretion states hereby produce
\begin{eqnarray}
 R_{jD}=\frac{E_j}{E_D}
 \label{EQN_RjD}
 \end{eqnarray}
 satisfying $R_{jD}<1$ \citep{van03} or $R_{jD}>1$ \citep{mck12}  representing a minor, respectively, major output in BPJ relative to the
 energy $E_D$ dissipated in the inner disk or torus. Summarizing the above, a black hole spin down phase satisfies
\begin{eqnarray}
R_j + R_k + R_D + R_S = 1,
\label{EQN_RS1}
\end{eqnarray}
where $R_D=E_D/E_r$ and $R_S$ denotes the fraction of spin energy dissipated unseen in the black hole event horizon, merely to increase the
Bekenstein-Hawking entropy. By (\ref{EQN_Rjr}-\ref{EQN_Rkr}), (\ref{EQN_RS1}) reduces to 
\begin{eqnarray}
R_D+R_S\simeq 1.
\label{EQN_RS2}
\end{eqnarray} 

Quite generally, the newly formed black hole evolves through the combined effect of mass-inflow and Maxwell horizon stress. The former may
derive from matter plunging in from the ISCO and injection by pair creation induced by the neutrino flux coming off the accretion
disk \citep{woo93}. The black hole tends to spin up (down) whenever the former is dominant (sub-dominant) over the latter according to 
a critical hyper-accretion rate \citep{glo14}. This may be probed by the secular evolution of GRB light curves, produced by BPJs 
induced by frame dragging about the spin axis of rotating black holes. 

\begin{table*}
\center{
\textbf{Table 1.} Sample of GRB-SNe. References refer to SNe except for GRB 070125. $E_*$ is expressed in units of $10^{51}$ erg.\\ 
\begin{tabular}{llccccrrrr}
\hline
GRB & Supernova				& $z$ & $E_\gamma$ & $E_{SN}$  & $\eta$ & $E_{r}/E_c$ & Ref.\\
\hline
980425		&Sn1998bw 	& 0.008 	        	& $<0.001$     	& 50	  	& 1	& 1.7		& 1\\
031203		&SN2003lw 	& 0.1055	        	& $<0.17$ 	 & 60  	&0.25& 10	& 2\\
060218 		&SN2006aj	& 0.033    	 	&$<0.04$ 		 & 2   	&0.25& 0.25	& 3 \\
100316D		&SN2006aj	& 0.0591		& 0.037-0.06	 & 10 	&0.25& 1.3	& 4\\
030329		&SN2003dh	& 0.1685		& 0.07-0.46       	& 40		&0.25& 5.3	& 5 \\
\hline
\hline
\end{tabular}
\label{TABLE_1}
}
\mbox{}\\\hskip0.08in
1. \cite{gal98}; 2. \cite{mal04}; 3. \cite{mas06,mod06,cam06,sol06,mir06,pia06,cob06b}; 
4. \cite{cho10,buf11}; 5. \cite{sta03,hjo03,mat03}; 
\end{table*}

We set out to model the two different evolutionary phases of black holes by hyper-accretion and interaction with matter at the ISCO, the first based on a new system of modified Bardeen equations. We next revisit a split horizon flux topology supporting $E_j$ and a major output to the ISCO and discuss the lifetime of black hole spin unifying long GRBs and SGRBEEs. Model predictions are confronted with the BATSE data and Table 1 with a discussion on implications for upcoming probes by gravitational wave detectors LIGO-Virgo and KAGRA.

\section{Bardeen accretion with outflows} 

\begin{figure}
\centerline{ \includegraphics[scale=0.8]{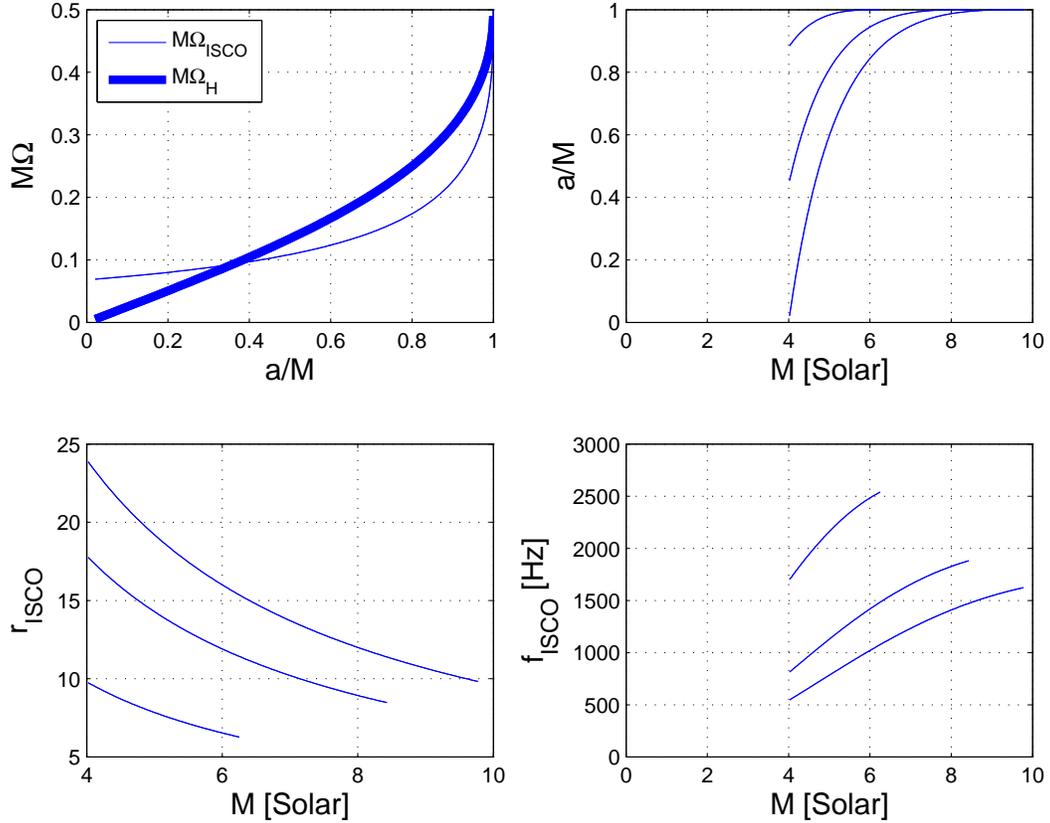} }
\caption{Shown is the surge of a black hole of initial mass $4M_\odot$ by hyper-accretion following
the Bardeen limit of zero outflows. In geometrical units, the left top panel shows the dimensionless product 
$M\Omega$ for the angular velocity $\Omega_H$ of the black hole and the angular velocity $\Omega_{ISCO}$
of test particles at the radius $r_{ISCO}$ (left bottom panel) of the Inner Most Stable Circular Orbit, 
both as a function of the dimensionless Kerr parameter $a/M$. The right top panel shows $a/M$ as a 
function of black hole mass. The top bottom panel shows the orbital frequency at $r_{ISCO}$.}
\label{fig1}
\end{figure}

In Bardeen accretion, the black hole of mass $M$ and angular momentum $J=aM$ evolves by matter plunging in from the ISCO \citep{bar70,kin06} with energy $e=\sqrt{1-2/3z}$ and angular momentum $j=(2M/3\sqrt{3}) (1+2\sqrt{3z-2}),$ where $z=r_{ISCO}/M$ of the radius of the ISCO in Boyer-Lindquist coordinates of the Kerr metric. Accretion is believed to be mediated by large eddy turbulent viscosity \citep{sha73}. If mediated turbulence, e.g., by the magneto-rotational instability \citep[MRI;][]{bal91,haw91}, magnetic flux forcefully advects onto the event horizon. This may lead to the formation of magnetized jet that, in the force-free limit, carries off an angular momentum flux $2L_j/\Omega_H$ to infinity at a luminosity $L_j$ \citep{bla77}. The black hole evolution hereby satisfies 
\begin{eqnarray}
\left\{\begin{array}{rll}
\dot{M} &= e\dot{m}-L_j,\\ \\
\dot{J} &= j\dot{m} - \frac{2L_j}{\Omega_H}.
\end{array}
\right.
\label{EQN_A}
\end{eqnarray}
The limit $L_j=0$ has the integral $zM^2=\mbox{const.}$ \citep{bar70}, shown in Fig. \ref{fig1}.
\begin{figure}
\centerline{ \includegraphics[scale=0.8]{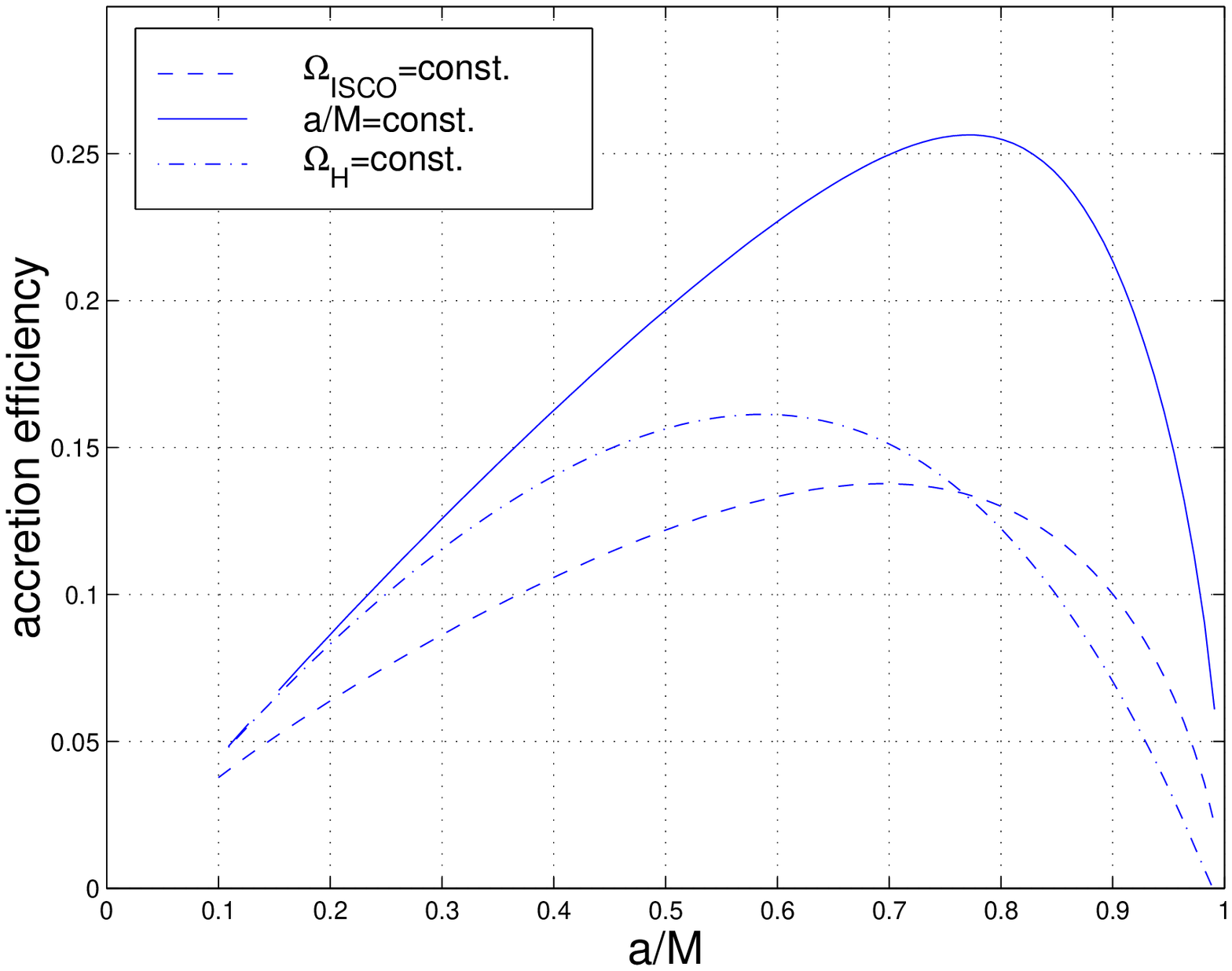} }
\caption{Modified Bardeen equations (\ref{EQN_A}) of black hole accretion with outflows allow
for solutions keeping selected properties of the black hole constant, here parameterized by
accretion efficiency in creating outflows. Shown are the critical lines 
of accretion efficiency in $L_j = \epsilon e \dot{m}$ for which $\Omega_{ISCO}$, $a/M$ or $\Omega_H$ are 
constant. At the lowest efficiencies, the results reduce to Bardeen accretion, for which these quantities
are naturally increasing. }
\label{fig2}
\end{figure}

For accretion of energy in magnetic flux as a function of the accretion rate $\dot{m}$, Fig. \ref{fig2} shows 
the limits of accretion efficiency $\epsilon$ in $L_j= \epsilon e \dot{m}$ as a function of $a/M$ with 
angular velocity $\Omega_{ISCO}=M^{-1}(z^\frac{3}{2}+a/M)$ at the ISCO. 
At lower (higher) efficiencies, they increase (decrease) in time. 

Some numerical experiments in steady state accretion flow point to an accretion efficiency 
$\epsilon \simeq 0.07(\hat{a}/(1+\sqrt{1-\hat{a}^2})^5$ \citep{mck05,kum08b}, i.e., $\epsilon \simeq 0.07\,(2M\Omega_H)^5$
with $2M\Omega_H=\tan(\lambda/2)$, $\hat{a}=\sin\lambda$. At these relatively moderate efficiencies, the black hole evolves essentially along a Bardeen trajectory with spin up in the angular velocity of the black hole up to a rapidly spinning black hole with $a/M\simeq 0.9$ 
as shown in Fig. \ref{fig2}. 

In the Bardeen limit $L_j=0$, \cite{tho74} points out that Bardeen accretion results in near-extremal Kerr black holes with a dimensionless spin parameter limited to $a/M\simeq 0.998$ $(\lambda=0.9597)$ due to radiation losses from the inner disk itself. Correspondingly, the black hole angular velocity and rotational energy are limited to 84.25\% and 92.49\% of their maximal values for a given $M$ ($2M\Omega_H = 0.8425$, $[\sin(\lambda/4)/\sin(\pi/8)]^2=0.9249$). This extremum defines the Thorne limit of Bardeen accretion. 

Spin down of black holes to moderate values of $a/M$ requires large efficiencies, e.g., from intermittency \citep{van99a,mck12,van14b}). 
If $L_j$ represents an opening outflow supported by horizon flux over the full hemisphere of the black hole, $E_j$ can then be a substantial 
fraction of $E_r$. For the prompt GRB phase, this is ruled out by $R_{jr}<<1$.

\section{Small $E_\gamma$ from split horizon flux}

Interactions of near-extremal black holes is facilitated by an equilibrium value of Carter's magnetic moment $\mu_H=QJ/M$ \citep{car68} in the lowest energy state. It preserves essentially maximal horizon flux in exposure to an external magnetic field from an inner accretion disk. A split of horizon flux into open outflows from a polar cap and a surrounding torus magnetosphere facilitates a minor ratio of $E_j$ to the total energy dissipation $E_D$ in the inner disk \citep{van03},
\begin{eqnarray}
R_{jD}<1,
\label{EQN_RjD}
\end{eqnarray}
in the presence of strong interactions with surrounding matter at the ISCO. Determined essentially by the Kerr metric, it obtains model light curves for the prompt GRB emission with essentially no free parameters, assuming a linear relationship between the luminosity in gamma-rays and in ultra-relativistic outflows from the black hole. The result can be confronted with light curves from BATSE.

Black holes of zero electric charge expel external magnetic flux according to \citep{wal74,dok86,tho86} $\Phi_H^0 = 4\pi M^2 \cos\lambda,$ where $\sin\lambda = a/M$ \citep{van99b}. This effect can be safely neglected in models of magnetic outflows from slowly spinning black holes \citep{bla77}. For extremal black holes ($\lambda = \pm \pi/2)$, interactions are preserved by an equilibrium $\mu_H^e$ upon acquiring a finite equilibrium electric charge \citep{wal74,ruf75}, that represents the minimum of ${\cal E} \simeq {Q^2}/{2r_H} - \mu_H B$ at $Q^e \simeq BJ(R_H/M)$ \citep{van01p} with associated dipole magnetic flux $\phi(Q)= 4\pi Q^e M \Omega_H$ \citep{wal74,dok86}. Here, $r_H$ denotes the radius of the event horizon in Boyer-Lindquist coordinates. A similar result obtains from a solution of the Grad-Shavranov equation \citep{lee01}. The total horizon flux is $\Phi_H^e = \Phi_H^0 + \phi(Q^e) \simeq 4\pi BM^2$, adopting $Q^e = 2BJ(r_H/M)$, $\mu_H^e=Q^eJ/M$ \citep{car68,coh73}, as an exact solution in the approximation of a Wald field \citep{wal74}. See further \cite{lee01}. {\em $\mu_H^e$ is 
particularly important to nearly extremal black holes that may form in the \cite{tho74} limit of \cite{bar70} accretion}. Since $Q^e\simeq BJ$ develops on a light crossing time scale of the black hole, equilibrium is preserved also in the presence of turbulence.

$\mu_H^e$ can support open flux out to infinity from polar caps in the limit of vanishing accretion, the remaining connected to
matter at the ISCO via a surrounding torus magnetosphere. This split is a different in topology from in earlier works \citep{ruf75,bla77}. It implies small values $R_{jD}$, as most of the black hole output is onto matter at the ISCO orbiting at $\Omega_{ISCO} = M^{-1}/(z^{3/2}+a/M)$, $z=r_{ISCO}/M$ \citep{bar72}. When $\Omega_H>\Omega_{ISCO}$, the black hole may spin down, corresponding to $a/M>0.36$; $a/M>0.4433$ to disks with outflows in magnetic winds \citep{van12}, at the sub-critical accretion rates of \cite{glo14}.

\begin{figure}[h]
\centerline{\includegraphics[scale=0.2]{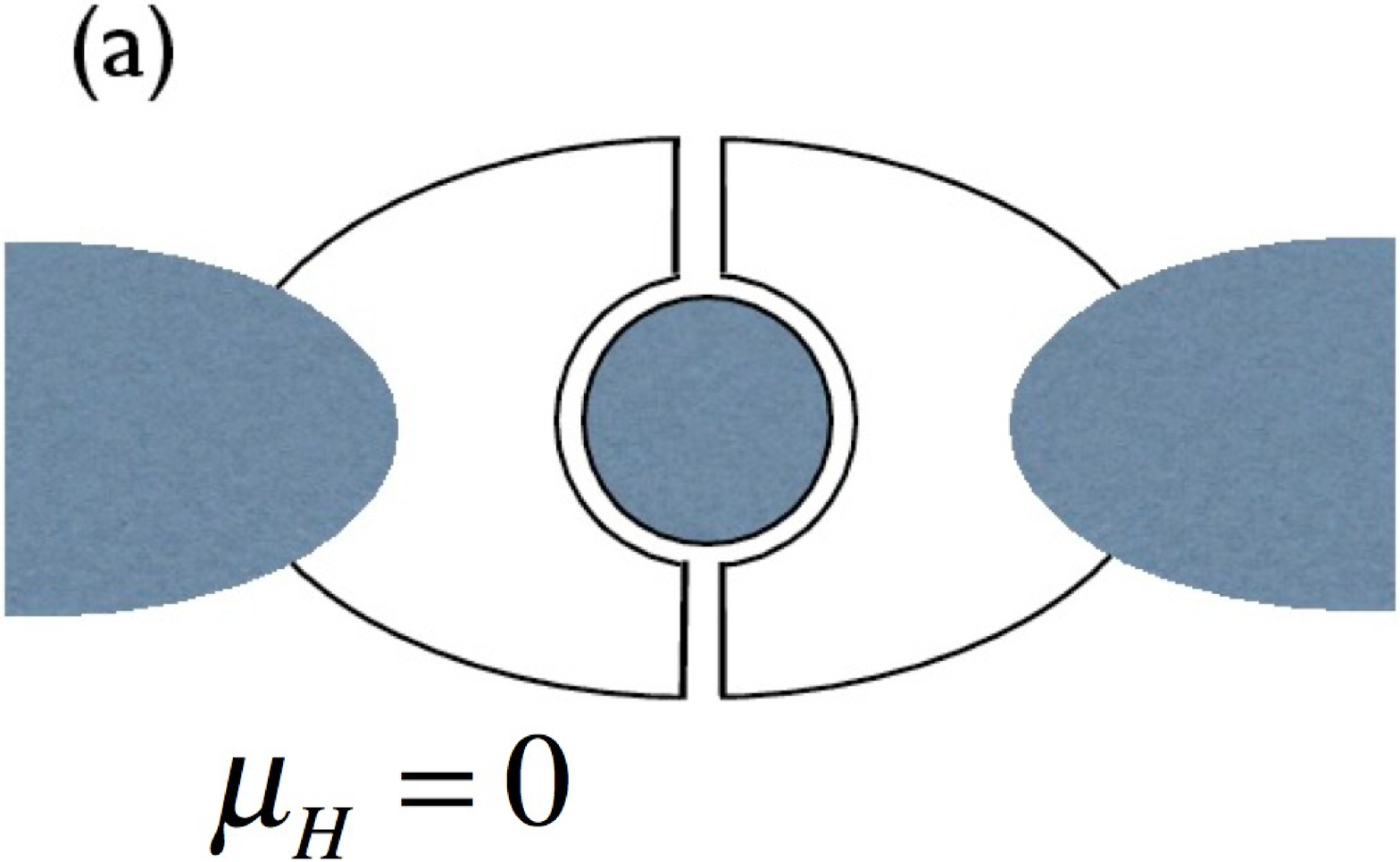}\includegraphics[scale=0.2]{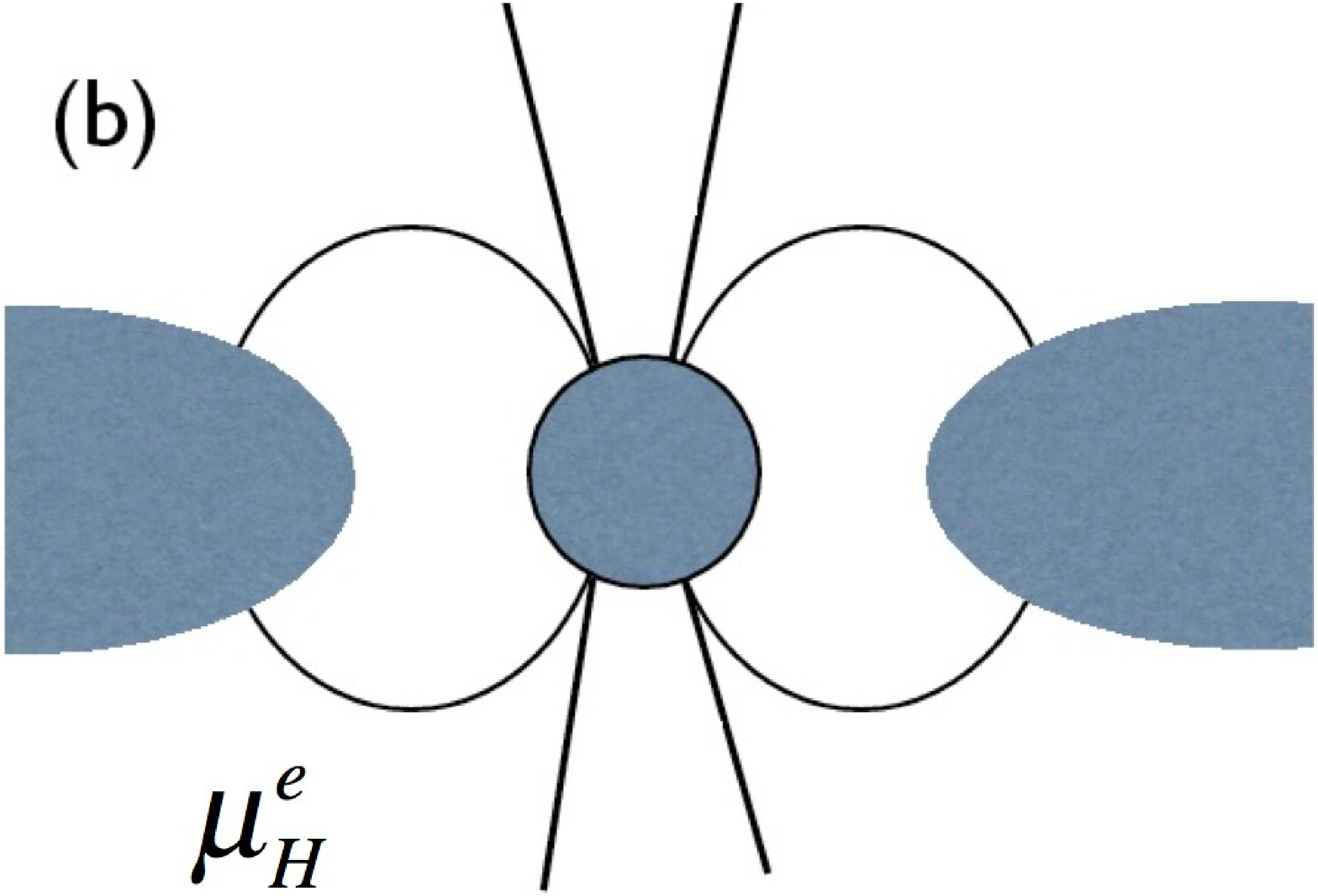}}
\caption{Sketch of an extremal black hole with magnetic pressure halted accretion for 
a vanishing Carter's magnetic moment (a) and the formation of an open magnetic flux tube supported by the equilibrium value $\mu_H^e$ (b).
In (a), the black hole does not evolve.  At sub-critical accretion rates in (b), frame dragging in the open flux may produce a BPJ while accretion from the ISCO is arrested by feedback from the black hole via an inner torus magnetosphere. Major dissipation ($D>0$) by forced MHD turbulence in matter at the ISCO implies $R_{jD}=E_j/D<<1$. With slip boundary conditions on the event horizon and no-slip boundary conditions on matter at the ISCO, the black hole spins down.}
\label{fig3}
\end{figure}

Fig. \ref{fig3} schematically illustrates the evolution of the black hole to its lowest energy state with associated induced dissipation in the inner disk. Magnetic pressure arrested disks inevitably turn to dissipative structures as in the suspended accretion model \citep{van99b,van01b} when baryon loading of the inner torus magnetosphere is sufficiently small, giving (\ref{EQN_RjD}). 

\begin{figure}[h]
\centerline{\includegraphics[scale=0.8]{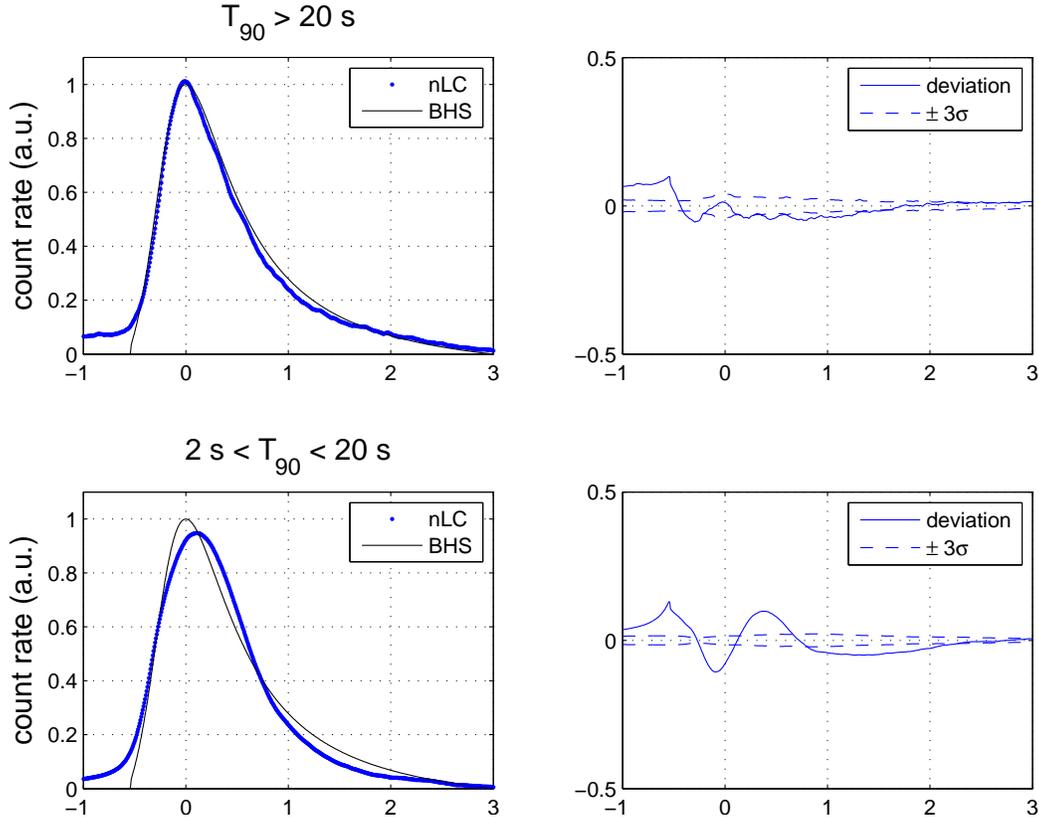}}
\caption{Normalized GRB light curves (nLC, thick lines) extracted from the BATSE catalog by matched filtering using model templates (thin lines) from rotating black holes losing angular momentum to matter at the ISCO according to (\ref{EQN_ode1}). (Adapted from \citep{van12}.)}
\label{fig4}
\end{figure}

\section{Near-extremal onset in BATSE} 

Alternatively to (\ref{EQN_A}) is evolution by Maxwell stresses alone \citep{van12}
\begin{eqnarray}
\left\{
\begin{array}{rll}
\dot{M} & = \Omega_T \dot{J},\\ \\
\dot{J} & =-\kappa e_k (\Omega_H-\Omega_T)
\end{array}
\right.
\label{EQN_ode1}
\end{eqnarray}
with $\kappa\simeq uM$, where $u\simeq 1/15$ is the ratio of total energy ${E}_{B,p}$ in poloidal magnetic field relative to the kinetic energy ${ E}_k$ of a torus at the ISCO. Roughly, the specific kinetic energy satisfies $e_k\simeq\frac{1}{2}v^2 e$, where $v=Mz\Omega_{ISCO}$. In contrast to (\ref{EQN_A}), (\ref{EQN_ode1}) posits a major interaction with matter at the ISCO contemporaneously with a minor output $L_j$ 
in a BPJ supported by $\mu_H^e$, even in the absence of accretion. It implies enhanced dissipation $D$ in a surrounding torus with $R_{jD}<<1$. According the black hole luminosity $L_H=-\dot{M}$ \citep{van99b} in (\ref{EQN_ode1}), {\em the lifetime of a black hole-torus system 
is the lifetime of black hole spin}, determined by the dimensionless mass $\sigma_T=M_T/M$ of the torus as \citep{van03}
\begin{eqnarray}
T_{spin} \simeq 30\,\mbox{s}\, \left(\frac{\sigma_T}{0.01}\right)^{-1}\left(\frac{z}{6}\right)^4\left(\frac{M}{7M_\odot}\right),
\end{eqnarray}
assuming the aforementioned value $u\simeq 1/15$. 

\begin{figure}[h]
\centerline{\includegraphics[width=160mm,height=70mm]{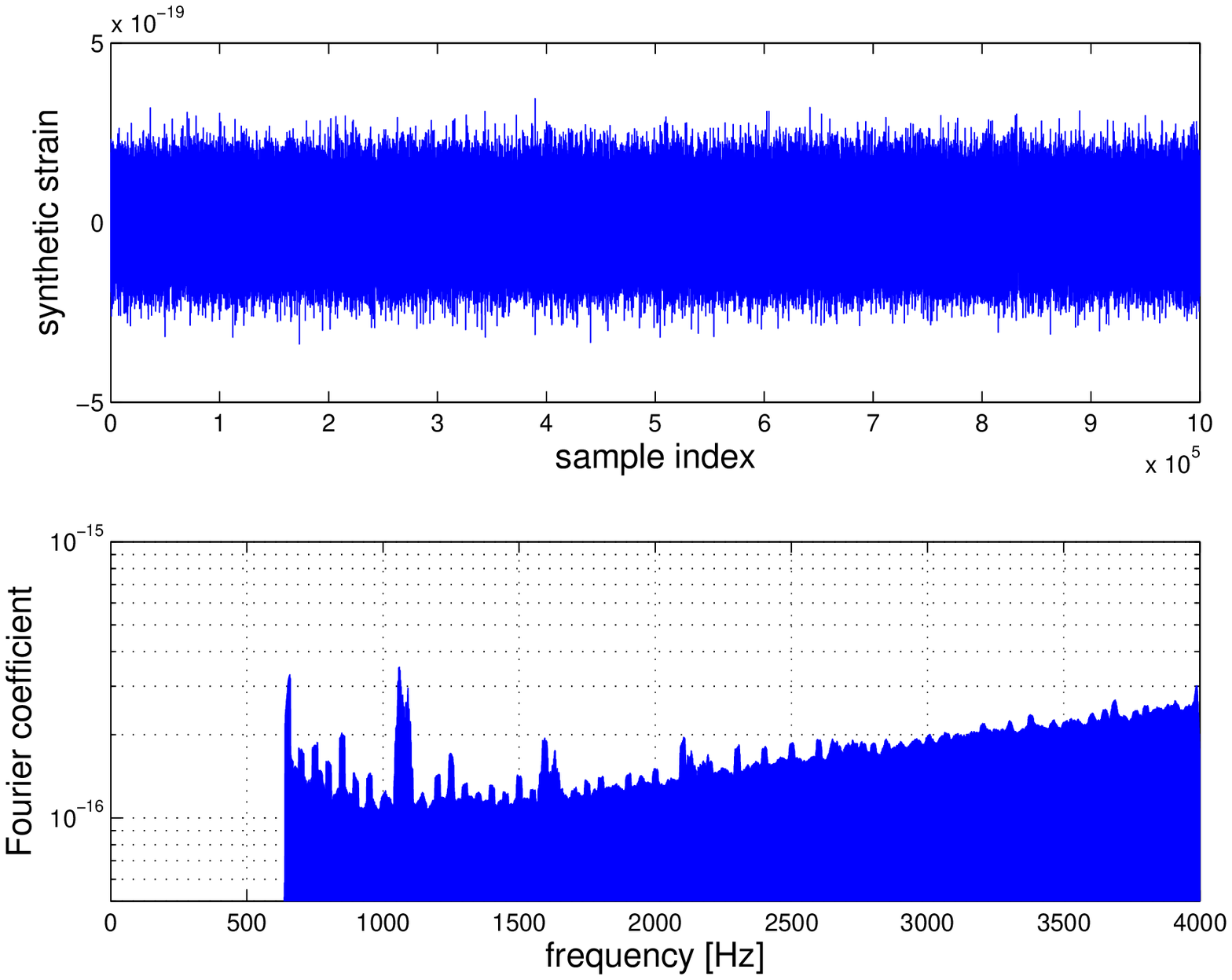}}
\caption{(Top panel.) Synthetic shot-noise representative for the high frequency detector strain-noise output of a laser interferometric gravitational wave detector. 
(Bottom panel.) Frequency spectrum of the synthetic shoot noise showing a characteristic increase with frequency.}
\label{fAPD_C0}
\end{figure}
\begin{figure}[h]
\centerline{\includegraphics[width=160mm,height=70mm]{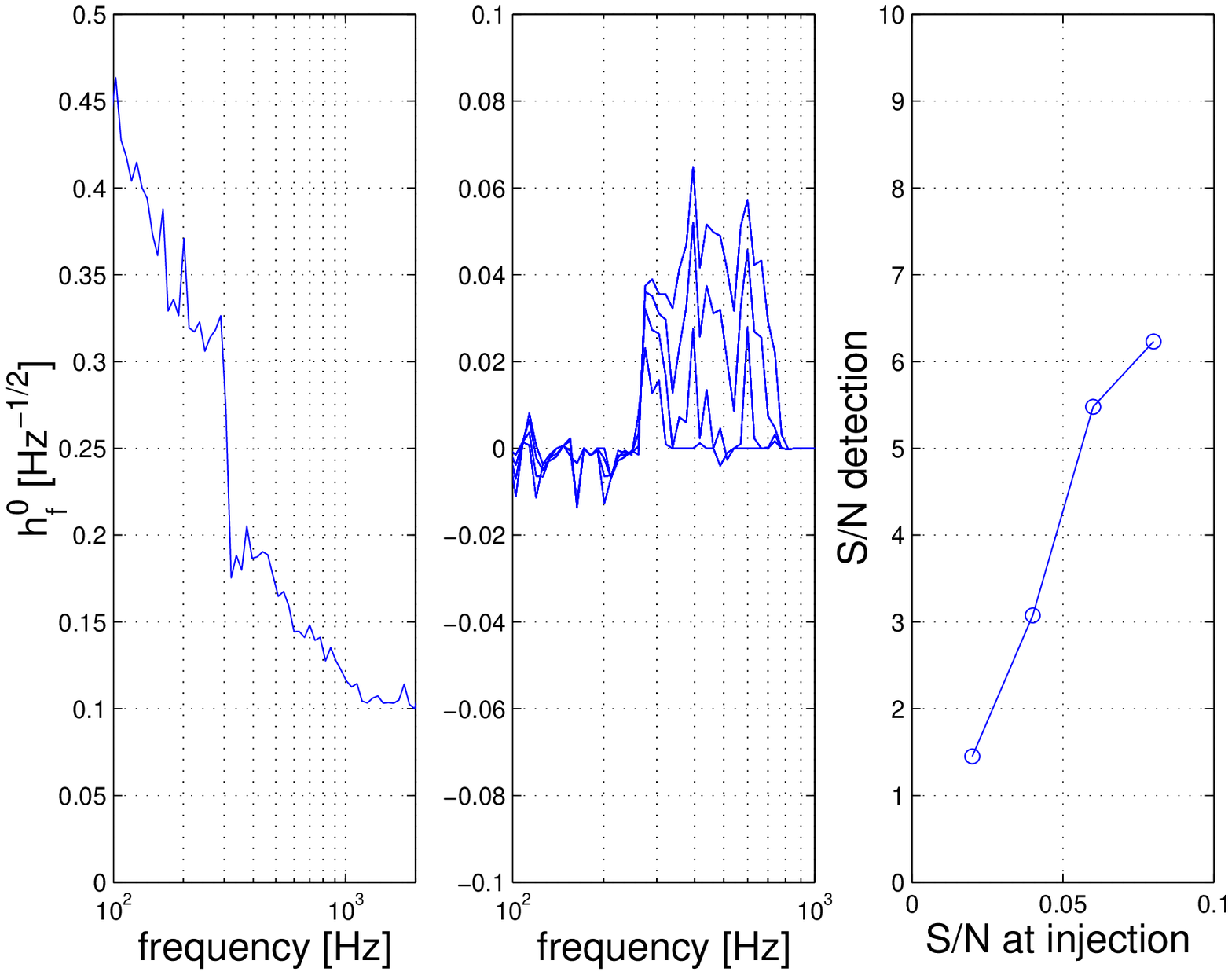}}
\caption{(Left panel.) Broadband spectrum of the synthetic shot-noise of Fig. \ref{fAPD_C0} produced by TSMF using a bank of 8.4 million chirp
templates. TMSF suppresses all constant frequency signals and detects only slowly changing frequencies, giving rise to the spectrum shown
distinct from the Fourier spectrum of Fig. \ref{fAPD_C0}. (Middle and right panels.) Source detection by identification of bumps in broadband spectra extracted following signal injection of a long duration model chirp by TSMF (25 slices, $\tau=1$ s, 8 million templates). The S/N in bumps is defined by the mean relative to the STD of $\Delta h_f$ over 100-1000 Hz. It obtains S/N=3.07 for injection at S/N $\simeq 0.04$, in agreement with the expectation 4.47 based on 25 slices and the late time frequency of $\sim$500 Hz at injection.}
\label{fAPD_C}
\end{figure}
In the split horizon flux topology, a model light curve of prompt GRB emission obtains from open outflows subtended by a half-opening angle $\theta_H$ on the event horizon. We consider $\theta_H=\theta_H(t)$ to be positively correlated to the ISCO with $L_j\propto \Omega_H^2$. In geometrical units, $L_j$ is dimensionless, i.e., $L_j\propto (r_{ISCO}\Omega_H)^2$ for such correlation to the ISCO. It obtains $L_j\propto z^2\Omega_H^2$ in a leading order Taylor series expansion in $z$. Integrating (\ref{EQN_ode1}) obtains a model light curve. For initially near-extremal Kerr black holes, Fig. \ref{fig4} shows a confrontation with a normalized light curves from BATSE. Events with $T_{90}> 20$ s show a particularly good match.

The results of \S2-3 point to favorable initial conditions for black hole spin-down after hyper-accretion has ceded, i.e.,
after a near-extremal state in spin. The resulting energy output onto matter at the ISCO may be self-regulated by feedback
through cooling in gravitational waves \citep{van12} and mass-injection \citep{glo14}. 

\section{Probing nearby CC-SNe for GWs} 

Based on this outlook, we propose surveying all nearby energetic core-collapse supernovae as targets of opportunity 
for probes by upcoming gravitational wave detectors \citep{heo15} in searches for long duration gravitational 
wave emission from accretion induced collapse of a neutron star \citep{mel14}, hyper-accreting black holes \citep{kob03,pir07} 
and non-axisymmetric motion at the ISCO \citep{van01p}.

To this end, we recently developed a Time Sliced Matched Filtering (TSMF, \cite{van11b,van14b}), that exploits phase coherence
in the anticipated emissions on intermediate time scales. Using banks of millions of templates, TSMF densely covers the frequency 
range and frequency time rate-of-change. In \cite{van14b}, the power of this method is demonstrated by identifying the 
broadband Kolmogorov spectra in light curves of long GRBs up to 1 kHz (in the observer's frame of reference), extracted
from BeppoSAX light curves with, on average, merely 1.26 photons in each bin of 500 $\mu$s. 

In the present discussion, we seek an application of this method to the strain detector output of gravitational wave detectors
probing hyper-energetic CC-SNe in the Local Universe. For these, the theoretical upper bound for the effective dimensionless strain
after ideal matched filtering satisfies \citep{van01a}
\begin{eqnarray}
h_{eff} \simeq \left(\frac{M_1}{D_{2}}\right) \left(\frac{E^{GW}_{-1}}{M_1}\right)^\frac{1}{2}\simeq 10^{-21},
\label{EQN_heff}
\end{eqnarray}
where $D=100D_{2}$ Mpc, $M=10 M_1 \,M_\odot$ and $E^{GW}=0.1 E^{GW}_{-1} M_\odot$ denotes the energy
output in gravitational waves.

Fig. \ref{fAPD_C0} shows synthetic broadband shot-noise typical for the high frequency gravitational strain output of  
LIGO-Virgo and KAGRA. Fig. \ref{fAPD_C} shows the spectrum extracted by our TSMF after application of a bank of over
8 million chirp templates of one second duration, representative for an observation when no signal is expected, i.e.,
away from the time-of-onset of a core-collapse supernova event. 
These chirp templates probe coherent features that slowly change
in frequency over an intermediate duration $\tau=1$ s. We then inject model signals, here of long duration broadband chirps
illustrative for gravitational wave emission from matter at the ISCO around a black hole spinning down, and calculate the
difference in spectra with and without injection. We define the S/N ratio of this difference by the mean 
over standard deviation in bumps as shown. Within a factor of two, TSMF recovers the theoretical limit of S/N in the 
proposed broadband detection method. TSMF hereby suffices for {\em source detection}, sacrificing all identification 
of detailed temporal signal behavior.

Following our injection experiment, the true effective strain in TSMF satisfies \citep{van11b}
\begin{eqnarray}
h_{eff}^\prime = h_{eff} \sqrt{\frac{\tau}{T_{90}}}\sim 10^{-22},
\label{EQN_heff2}
\end{eqnarray}
where the duration $T_{90}$ of the burst is here identified with the durations of tens
of seconds of long GRBs \citep{van11b}. The results point to a sensitivity distance of about 100 Mpc for
advanced LIGO at high laser power operation \citep{lig10}. 
For the proposed source detection by adding results of each chirp template matched filtering in the frequency domain as 
shown in Fig. \ref{fAPD_C}, the true effective strain obtained will be between the limits (\ref{EQN_heff}) and (\ref{EQN_heff2}).  

\section{Conclusions and outlook}

A combination of phenomenological evidence points to a common origin of GRBs from rotating black holes with spin down during the phase of soft EE, rather than spin down of rapidly rotating neutron stars or magnetars. In particular, we mention 
\begin{itemize}
\item the Amati-correlation common to soft EE from SGRBEE, LGRBN and normal LGRBs \citep{van14a}. Since the merger origin of the former almost certainly produce black hole inner engines, LGRBs hereby derive from the same;
\item no bump in the smooth high frequency extension of the Kolmogorov spectrum up to a few kHz in the comoving frame of long GRBs in {\em BeppoSax} light curves of bright LGRBs, indicating no trace of proto-pulsars \citep{van14b}; 
\item spin down in matched filtering analysis of the BATSE catalogue, showing remarkably consistency for spin down against high density matter at the ISCO for events with relatively long duration events $T_{90} > 20$ s (Fig. \ref{fig4});
\item the lifetime of rotating black holes provides natural secular time scale of tens of seconds in the merger scenario of SGRBs involving rotating black holes, allowing soft extended emissions in SGRBEEs and LGRBNe of similar durations in the absence of long duration hyper-accretion;
\item the existence of hyper-energetic events that defy the energy limits of rotating neutron stars or magnetars (Table 1), unless considerable energy also derives from disk winds emanating from hyper-accretion flows onto (proto-)neutron stars; 
\end{itemize}

The third item in particular points to near-extremal black holes at the onset of long duration prompt emission in LGRBs. We identify these near-extremal states with the result of hyper-accretion in a precursor phase followed by prompt GRB emission during spin-down. Most of the black hole spin 
energy is herein dissipated and radiated off in gravitational waves in the surrounding high density matter, i.e., $R_{jD}<1$ with accompanying
small values of the first two calorimetric ratios $R_{jr}$ and $R_{kr}$. The lifetime of black hole spin herein gives a natural unification of long durations in LGRBs from core-collapse events and SGRBEEs from mergers. In exhausting $E_r$, major emissions unseen are anticipated in gravitational waves and MeV neutrinos in light of the remarkably small ratios $R_{jr}$ and $R_{kr}$. Our anticipated small ratio $R_{jD}$ points to a major output in gravitational wave emission that may be probed by LIGO-Virgo and KAGRA in nearby hyper-energetic CC-SNe, possibly identified in dedicated surveys \citep{heo15}. Detection of this output by the proposed TSMF promises true calorimetry on the inner engine of the most energetic transients in the Universe. 

Gravitational wave detection offers a unique window to the detailed observation of formation and evolution of black hole inner engines in CC-SNe.
It may solve a number of open questions that are otherwise difficult to constrain with any certainty. Here, we mention just two constraints on the inner engine of GRBs that are challenging to constrain from existing electromagnetic observations.  

For LGRBs, successful break-out puts some constraint on relativistic outflows from the inner engine. In the limit of relativistic hydrodynamics, continuous outflows lasting at least some ten seconds are required \citep{bro11,laz12}. In the limit of relativistic MHD, this requirement is relaxed by intermittency producing long-lived magnetic ejecta \citep{van14b}. Yet, a minimum activity time of about ten seconds may be inferred from observations \citep{van12,bro14}, suggesting that early dissipation to an effectively hydrodynamic jet might indeed have occurred \citep{lev13}. Whether a similar constraint carries over to the limit of Poynting flux dominated outflows \citep{bro15} is unclear. 

The present model for prompt GRB emission from initially near-extremal black holes opens a window to precursor outflows during the prior surge during hyper-accretion along a modified Bardeen trajectory. The light curve thus produced would asymptote to a relatively flat plateau, which is not consistent with normalized light curves extracted from the BATSE catalogue \citep[Fig. 3 in][]{van09a}. Relatively few GRBs, both short and long, give evidence of such precursors \citep[e.g.][and references therein]{tro10}, that might be accounted for by the activity of the central engine. Conceivably, additional structure of outflows from hyper-accreting black hole disk systems is different from black hole torus systems in suspended accretion, e.g., in luminosity and the degree of collimation. If so, the paucity of precursors could be an observational selection effect. Furthermore, the transition to the proposed suspended accretion state should be prompt, at the point when hyper-accretion drops below the \cite{glo14} critical value for hyper-accretion along the proposed modified Bardeen trajectory. Below this critical value, feedback from the black hole sets in, bringing hyper-accretion to a halt leaving an essentially Poynting flux dominated torus magnetosphere around the black hole \citep{van99b}. At this point, a relatively minor amount of mass in the ensuing torus suffices, to support the required super-strong magnetic fields \citep{van03}. 

A direct observation on the activity of the inner engine in CC-SNe by gravitational wave emissions promises to identify the detailed activity and transient behavior of the inner engine by identification of specific gravitational wave features, in temporal behavior and gravitational wave spectra. For the NS-NS and NS-BH mergers, it would be interesting to identify them by different black hole masses and spins, otherwise difficult to pursue by existing electromagnetic observations. For this reason, we emphasize the need for essentially maximally sensitive broadband detection algorithms by exploiting modern high performance computing. 

\mbox{}\\
{\bf ACKNOWLEDGMENTS.} The author gratefully acknowledges support of the Faculty Research fund of Sejong University in 2012 and 
constructive comments from the referees which considerably improved the readability of the manuscript.

\end{document}